\documentclass[12pt]{article}
\usepackage{jcapmod}
\usepackage{rotating}
\usepackage{amsmath}
\usepackage{mathtools}
\usepackage{amsthm}
\usepackage{amsfonts}
\usepackage{graphicx}
\usepackage{psfrag}
\usepackage{amssymb}
\usepackage{slashed}
\usepackage{lscape}
\usepackage{url}
\usepackage{bm}
\usepackage{subfig}
\usepackage{epsfig}
\usepackage{tikz}
\usepackage{tikz-cd}

\def\ga{\mathrel{\raise.3ex\hbox{$>$\kern-.75em\lower1ex\hbox{$\sim$}}}}
\def\la{\mathrel{\raise.3ex\hbox{$<$\kern-.75em\lower1ex\hbox{$\sim$}}}}

\def\beq{\begin{equation}}
\def\eeq{\end{equation}}
\def\bea{\begin{eqnarray}}
\def\eea{\end{eqnarray}}
\def\beqar{\begin{eqnarray}}
\def\eeqar{\end{eqnarray}}
\def\avg#1{\langle #1 \rangle}

\def\iso#1#2{\mbox{${}^{#2}{\rm #1}$}}
\def\he#1{\iso{He}{#1}}
\def\li#1{\iso{Li}{#1}}
\def\be#1{\iso{Be}{#1}}
\def\bor#1{\iso{B}{#1}}

\def\pfrac#1#2{\left( \frac{#1}{#2} \right)}

\begin{document}

\begin{flushright}
UMN--TH--4118/22, FTPI--MINN--22/09   \\
April 2022
\end{flushright}

\title{Implications of the Non-Observation of \boldmath ${}^{6}{\rm Li}$ in Halo Stars for the Primordial ${}^{7}{\rm Li}$ Problem}

\author[a]{Brian D. Fields}
\author[b]{Keith A. Olive}

\affiliation[a]{Illinois Center for the Advanced Study of the Universe, and
Departments of Astronomy and of Physics, University of Illinois,
Urbana, IL 61801}
\affiliation[b]{William I. Fine Theoretical Physics Institute, School of
 Physics and Astronomy, University of Minnesota, inneapolis, MN 55455}

\abstract{
The primordial Lithium Problem is intimately connected to the assumption that \li7 observed in 
the atmospheres of metal-poor halo stars in fact retains its primordial abundance, which lies significantly below the predictions of standard big-bang nucleosynthesis.
Two key lines of evidence have argued
that these stars have not significantly depleted their initial (mostly primordial) \li7:
i) the lack of dispersion in Li abundance
measurements at 
low metallicity (and high surface temperature); and ii) the detection of the more fragile
\li6 isotope in at least two halo stars. The purported \li6 detections were in good agreement with predictions from
cosmic-ray nucleosynthesis which is responsible for the origin of \li6.
This concordance left little room for \li6 depletion, and
the apparent \li6 survival
implied that \li7 largely evaded
destruction, because stellar interiors destroy
\li6 more vigorously then than \li7.
Recent (re)-observations of halo
stars challenge the evidence against \li7 depletion:
i) lithium elemental abundances now show significant
dispersion, and ii) sensitive \li6 searches
now yield no definitive detections,
revealing only firm upper limits to the \li6/\li7 ratio.
We discuss the consequences of these \li6 non-detections on the primordial \li7 Problem, Galactic cosmic-ray nucleosynthesis, and the question of differential depletion of Li in stars. 
The tight new \li6 upper limits generally fall far below the predictions of
cosmic-ray nucleosynthesis, implying that substantial \li6 depletion has occurred--by factors up to 50.  
We show that in stars with \li6 limits and thus lower bounds on \li6 depletion,
an equal amount of \li7 depletion is more than sufficient to
resolve the primordial \li7 Problem.  In fact, this picture is consistent
with stellar models in which \li7 is less depleted than \li6, and strengthen the case
that the Lithium Problem has an astrophysical solution. We conclude by
suggesting future observations that could test these ideas.
}

\maketitle

\section{Introduction}
There was a great deal of excitement when the first observations of lithium in low metallicity halo stars appeared.  The Li abundances were uniform and
independent of metallicity\footnote{Here the ``metallicity'' 
$[{\rm Fe/H}]=\log_{10}[{\rm (Fe/H)/(Fe/H)_\odot}]$ is the iron abundance expressed a log relative to its solar value.} for $[{\rm Fe/H}]  < -1.5$ \cite{spites}, implying a
pre-stellar origin for Li. Indeed,
the observed abundance coincided well with the prediction of standard Big Bang Nucleosynthesis (BBN) \cite{ytsso}. Subsequently, there have been many observations of Li in stars with $-3 <$ [Fe/H] $< -1.5$
\cite{rnb,ryan2000,melendez2004,bonifacio2007,hos}
defining what is commonly known as the Spite plateau.
Furthermore, the only dispersion seen in this plateau seemed consistent with observational uncertainty
\cite{rnb}.
Only the primordial species D, \he3, \he4, and \li7 produced in BBN
are expected to show little or no variation at very low metallicity, 
potentially justifying the association of the \li7 in the plateau with the primordial abundance of \li7. 

So long as the uncertainties in the baryon-to-photon ratio, $\eta$, and primordial abundances were sufficiently large,
there was concordance between light-element abundances as observed versus
the predictions from BBN theory which depend on $\eta$ \cite{bbn}.
This picture began to break down when higher precision data on D/H became available \cite{reliable}. This D/H (confirmed with high precision by several more observations \cite{otherD}) indicated a relatively large baryon-to-photon ratio $\eta \sim 6 \times 10^{-10}$ which would predict 
a \li7/H abundance in excess of the plateau value (discussed in more detail below in \S\ref{liprob}).
Shortly thereafter, measurements of the cosmic microwave background (CMB) by {\em WMAP} \cite{wmap1} and confirmed by {\em Planck} \cite{Planck2015,Planck2018} resulted in a very precise determination of $\eta = (6.12 \pm 0.04) \times 10^{-10}$, in excellent agreement with BBN calculations and the observational determination of D/H and \he4 \cite{cfo3,CFOY,FOYY}
exacerbating the \li7 Problem \cite{cfo5}.   The current primordial \li7 abundance from our group is \cite{Yeh2021}
\begin{equation}
\label{eq:Lip}
\left. \frac{\li7}{\rm H} \right|_{\rm BBN+CMB} = (4.94 \pm 0.72) \times 10^{-10} \, ,
\end{equation}
in good agreement with other recent work by the Paris group \cite{Pitrou2021} ((5.46 $ \pm 0.22) \times 10^{-10}$) and Naples group \cite{Parthenope} (4.69 $\times 10^{-10}$).
These abundances should be compared with the observational determination  \cite{ryan2000,sbordone2010}
\begin{equation}
\label{eq:LiO}
\left. \frac{\li7}{\rm H} \right|_{\rm obs} = (1.6 \pm 0.3) \times 10^{-10}  \, ,
\end{equation}
which falls a factor $\sim 3$ and over $4\sigma$ below the prediction in eq.~(\ref{eq:Lip}).
This deficit comprises the ``Primordial \li7 Problem,'' or simply the Lithium Problem.

The lack of dispersion in the \li7 abundance data
has often been used as an argument against the {\em in situ} depletion of \li7 in halo stars. For example, at higher metallicity or in stars with lower surface temperatures,
convective depletion is expected and is evidenced by the significant dispersion in the \li7 abundance in these stars. Furthermore, any depletion in \li7 should coincide with at least as much depletion of \li6 \cite{BS}. Therefore, the initial observations of \li6 in two halo stars \cite{sln,sln2}
was particularly important, as they further confirmed the lack of 
significant depletion, 
and strengthened the case for the primordial status of the Spite plateau.
Indeed the \li6 abundance in these two low-metallicity ([Fe/H] $\approx -2$) stars 
was entirely consistent with the expected \li6 production in Galactic cosmic-ray
nucleosynthesis (GCRN) \cite{FdO4,Vangioni99}, and therefore provided 
evidence against depletion in these stars.

The lithium isotopic ratio had been reportedly measured several times in HD 84937. In \cite{sln}, results from the observations of two stars, HD 84937 and HD 19445 were presented. In the former (with [Fe/H] = -2.4 and $T = 6090$ K), an isotopic ratio $\li6/\li7 = 0.05 \pm 0.02$ was determined, whereas in the latter (with [Fe/H] = -2.2 and $ T = 5820$ K) only an upper limit \li6/Li $ < 0.02$ could be established, consistent with the idea that depletion occurs in the cooler stars. Other positive determinations in HD 84937 were reported in \cite{sln2,otherli6}.  
The weighted average of the available measurements yielded \li6/Li $= 0.054 \pm 0.011$ at [Fe/H] $\simeq -2.3$. Subsequently, an additional positive detection in BD 26$^\circ$ 3578 with \li6/Li $= 0.05 \pm 0.03$,
at about the same metallicity was reported in \cite{sln2} and a weak detection in G271-161
with  \li6/Li $= 0.02 \pm 0.01$ \cite{naho}~\footnote{For a more complete summary of the early observations of \li6, see \cite{FdO4}.}. 

Several years after these initial measurements, there was a report of ($\ge 2 \sigma$) \li6  detections in 9 out of 24 stars observed \cite{asplund} over a metallicity range of -1.25 $<$ [Fe/H] $<$ -2.74.
But the real surprise in this work was the uniformity of the \li6/\li7 ratio as a function of metallicity.  At these metallicities, the \li7 abundance is primarily of BBN origin, and so is essentially constant
with [Fe/H]. 
\li6, on the other hand, is predominantly produced in GCRN \cite{crnuke,sfosw,Fields:1994kn,rklr,FdO3,FdO4,Vangioni99} and grows monotonically with [Fe/H] from its near negligible BBN abundance \cite{tsof,Vangioni99,coc3}. 
As a result, we expect that the isotopic ratio
should also increase with metallicity rather than form
an apparent \li6 `plateau'. Furthermore, 
these observational determinations imply that
\li6 was in fact depleted at metallicities 
$\gtrsim -2$, as cosmic-ray production is expected to exceed this `plateau' value.
Similarly, the determined abundance at lower metallicity necessitated enhanced production mechanisms such as Pop III production \cite{rvo}, flares \cite{flare}, or late-decaying particles \cite{decay}.

In a detailed study of HD 74000, it was found \cite{cayrel} that convective processes generate an excess of absorption in the red wing of the \li7 absorption spectrum that could be interpreted as the presence of \li6. 
The existence of the \li6 `plateau' was called into question. Indeed, it was estimated \cite{cayrel2} that the \li6/\li7 ratios determined in \cite{asplund} should be reduced by 0.015, thus reducing the number of detections from 9 to 4 (or less). This conclusion was reaffirmed  in \cite{GP} which
examined 5 stars finding no significant detections. One of the stars in this study
was BD 26$^\circ$ 3578 with an isotopic ratio of only 0.004 $\pm$ 0.028 (a ratio of 0.01 $\pm$ 0.013 was found in \cite{asplund} for the same star. 

The reliability of the \li6 `plateau' was further called into question with new analysis employing 3D NLTE modelling \cite{lind}. Four stars including HD 84937 were studied with no significant detections.  The newly determined isotopic ratio in HD 84937 was found to be be 0.011 $\pm$ 0.010 $\pm$ 0.011, i.e., consistent with no \li6. A recent study of a very metal poor star with [Fe/H] = -3.7, also found no detectable \li6, though this result might be expected in standard cosmic-ray models. 

The most recent nail in the coffin for the \li6 `plateau' came from new observations using the VLT/ESPRESSO spectrograph \cite{wang}. Three stars including HD 84937 were observed
and only upper limits to the \li6/\li7 ratio could be obtained. In the case of HD 84937,
where a ratio of 0.05 is in fact quite consistent with standard cosmic-ray nucleosynthesis \cite{FdO4}, a $2 \sigma$ upper limit of 0.007 was obtained.
As this result may now directly call for the 
stellar depletion of \li6, one can also call into question the that \li7 in halo
stars retains its initial primordial value.

Indeed, the durability of the \li7 plateau
has also come into question.  The first indications of a departure from a \li7 plateau
extending towards zero metallicity showed
both a correlation of \li7 with metallicity
and significant dispersion in the data
with metallicity [Fe/H] $\lesssim$ -2.7 \cite{aoki2009,sbordone2010}. In another study \cite{Bon15},
no ultra metal-poor star was found with a plateau value for \li7. Significant dispersion at low metallicity is also seen in more recent work \cite{Bon18,Aguado:2019egq,francois,Aguado:2020kki,pinto}. To wit, very recent observations \cite{mucciarelli22} of metal-poor red giant branch stars indicate a plateau, but at \li7/H $\approx 10^{-11}$, far below either the BBN abundance or even the initial Spite plateau. 

It nevertheless remains true that the \li7 plateau 
remains intact as an upper limit to the \li7 abundance in metal-poor stars. But there seems to be abundant evidence for dispersion and low(er) \li7 abundances when [Fe/H] $\lesssim$ -2.7. This combined with the recent lack of evidence of \li6 detection, may be pointing to stellar depletion processes \cite{dep,dep6,newdep} as a solution to the mismatch between the BBN prediction for primordial \li7/H and the upper envelope of the the \li7 observation, i.e., the \li7 plateau. 

In what follows, we will concentrate on the current state of the \li6 observations vis a vis predictions from GCRN. Our conclusion that \li6 destruction is likely to have occurred is consistent with the apparent
destruction of \li7 leading to the observed dispersion.  To this end, we briefly review 
the current status of the \li7 Problem in \S\ref{liprob}. In \S\ref{gcrn},
we take a fresh look at the predictions of GCRN of \li6 in comparison with recent observations. 
We discuss the implications of the non-observation of \li6 for stellar depletion and \li7 in \S\ref{nonobs}.
Further discussion and our conclusions are given in \S\ref{conc}.

\section{The Primordial \li7 Problem}
\label{liprob}

Stated very simply, the \li7 Problem is lack of agreement between the BBN prediction for \li7/H as given for example in Eq.~(\ref{eq:Lip}), and the observationally determined value given in Eq.~(\ref{eq:LiO}) \cite{fieldsliprob}. The BBN prediction for \li7 is a sensitive function of the baryon-to-photon ratio, $\eta$,
or equivalently the present baryon density parameter $\Omega_{\rm b} h^2 \propto \eta$ \cite{ytsso}. At relatively low values of $\eta$ ($< 3 \times 10^{-10}$), the BBN production of \li7 is a decreasing function of $\eta$. However at larger values of $\eta$, the \he3 ($\alpha, \gamma$)\be7 reaction dominates over $t(\alpha, \gamma)\li7$ and the primordial production of \be7 (which later decays to \li7) increases with $\eta$. 

As noted earlier, as long as the uncertainty in the other BBN element abundances and $\eta$ were sufficiently large,
concordance with \li7 observations was possible. However, 
in standard BBN, this requires $\eta \lesssim 4 \times 10^{-10}$ and D/H $\gtrsim 5 \times 10^{-5}$. Both values are strongly excluded by CMB measurements and D/H observations in quasar absorption systems. As a consequence,
attention was turned to 1) BBN reaction rates, 2) post-BBN processing of \li7, 3) {\em in situ} depletion of \li7 in stars. 

Because of the strong sensitivity of the \li7 abundance to the \he3 ($\alpha, \gamma$)\be7 rate (its log sensitivity is 0.964, i.e., it is nearly proportional to this rate \cite{FOYY}), attention turned to the experimental measurement of this cross-section \cite{he3alpha}.   
A subsequent analysis by Cyburt and Davids \cite{cybdav}
produced a model independent fit with significantly smaller uncertainties using a
Markov Chain Monte Carlo algorithm. However, the resulting
BBN \li7 abundance increased when the new rates were employed \cite{cfo5}. In fact it is not surprising that
attempts at resolving the Li Problem by re-examination of the standard BBN rates failed \cite{cfo4,coc3,boyd}
as these are the same rates operating in the Sun and confirmed by solar neutrino flux observations. 
Other rates such as $\be7(n,p)\li7$ \cite{damone2018,Iwasa:2021lse},
$\be7(n,\alpha)\he4$ \cite{Barbagallo2016,Hou2015,Kawabata2017,Lamia2017},
$\be7(d,p\alpha)\he4$ \cite{Rijal2018},
$\be7(\alpha, \gamma)\iso{C}{11}$ \cite{hartos2018}, 
$\li7(d,n\alpha)\he4$ \cite{Hou2021} have been recently been (re)measured
(or re-evaluated \cite{singh}), though none of these rates make more than a marginal change in the final \li7 abundance.

A potentially more interesting possibility is that a reaction thought to be 
unimportant could contain an undiscovered {\em resonance}
which could boost its cross section enormously,
analogously to the celebrated Hoyle \iso{C}{12} resonance
that dominates the $3\alpha \rightarrow \iso{C}{12}$ rate
\cite{Hoyle}. Examples of 
possible candidate resonances include:
$\be7 + d \rightarrow \iso{B}{9}^*$,
$\be7+ t  \rightarrow \iso{B}{10}^*$
and $\be7 + \he3 \rightarrow \iso{C}{10}^*$,
with various possible exit channels
\cite{cp,chfo,brog}. 
However, measurements in $\be7(d,d)\be7$ \cite{OMalley},
$\iso{Be}{9}(\he3,t)\iso{B}{9}$ 
\cite{Kirsebom},
and an $R$-matrix analysis of \iso{B}{9} \cite{paris}
all rule out a $\iso{B}{9}$ resonance.
Similarly,
 \iso{C}{10} data rule out the needed resonance in $\iso{C}{10}$ \cite{Hammache}.
The ``nuclear option'' to the Lithium
Problem is essentially excluded.

The post-BBN processing of \li7 generally involves physics beyond the Standard Model and is hence more speculative. 
The possibility that a long-lived particle decays during or after BBN, injecting electromagnetic and/or hadronic energy
into the Universe altering the BBN-produced abundances has been well-studied \cite{jed,decay2,ceflos1.5,serp}. 
However, these decays which may destroy \li7 (still \be7 at this time), generally also affect the abundances of D and \he4 \cite{dli,opvs}.
For example, the disruption of \he4 creates 
both D and neutrons as fragments.  These neutrons,
as well as nonthermal neutrons produced
in the case of hadronic decays,
then lead to mass-7 destruction
via $\be7(n,p)\li7(p,\alpha)\he4$,  driving D/H to higher values.  
Other non-standard model attempts include axion cooling \cite{sik} and variations of fundamental constants \cite{dfw,cnouv}. 
 However, the new very precise D/H measurements
\cite{otherD} dramatically reduce the possibility of  perturbing the BBN abundances (for which there is excellent agreement with D/H observations) and severely
challenge most new-physics solutions to the 
Lithium Problem.

\section{Galactic Cosmic-Ray Nucleosynthesis}
\label{gcrn}

A third class of potential solutions to the \li7 Problem is the {\em in situ} destruction of \li7 in halo stars. We will argue below that an upper limit to the depletion of \li7 can be obtained from \li6 observational upper limits. 
To infer lithium depletion from the \li6 observations
requires us to describe lithium isotope evolution with metallicity,
which determines the initial abundance in halo stars.
That is, we must have a model for the Galactic chemical evolution of the lithium
isotopes.  In this section we present results for a simple one-zone, closed-box model
for GCRN, based on our earlier work 
\cite{FdO3,FdO4}.  Our aim is to illustrate 
the procedure to infer depletion, using
a typical LiBeB chemical evolution
model.  Other GCRN models are certainly possible;
these will give different quantitative results, but our basic qualitative features and conclusions are robust.

\subsection{Cosmic-Ray Production of LiBeB}

The cosmic-ray production of LiBeB nuclides in the interstellar medium (ISM) includes two processes.
For the low-metallicity stars of interest,
the most important cosmic-ray source of lithium
is the {\em fusion} of $\alpha$ particles:
$\he4_{\rm CR} + \he4_{\rm ISM} \rightarrow \li{6,7} + \cdots$.
This process is selective, in that it only makes the \li{6,7} isotopes
and not Be or B.  
In the young Galaxy, $\alpha+\alpha$ fusion dominates because both the target and projectile
\he4 nuclei are abundant, having arisen
in BBN.
Thus, from the onset of cosmic-ray acceleration in the Universe,
this mechanism operates to produce lithium.

The more ``democratic'' mechanism for cosmic-ray nucleosynthesis
is {\em spallation}--the fragmentation of interstellar target nuclei heavier than
the  $\ell \in (\li6,\li7,\be9,\iso{B}{10},\iso{B}{11})$ isotopes of interest.  
That is, reactions such as
$p_{\rm CR} + \iso{O}{16}_{\rm ISM} \rightarrow \ell + \cdots$
can produce {\em all} LiBeB species, in 
ratios determined
by reaction branchings and cosmic-ray spectra.
The same reactions with ``inverse'' kinematics,
e.g., 
$\iso{O}{16}_{\rm CR}+p_{\rm ISM}$, are also included in our model,
as are losses of the LiBeB products which
have high energies and escape the Galaxy before being stopped.
Our model also includes as targets $\iso{C}{12}$ and $\iso{N}{14}$.
Recent work has argued for inclusion of even heavier and less abundant
targets up to $\iso{Fe}{56}$ \cite{Maurin:2022irz};
we neglect those contributions here which we expect to be small
and which will not affect our conclusion.

Note that spallation reactions require the existence of heavy nuclei 
that must first be produced by stars.
Consequently, production rates are low at early times
and low Galactic metallicities, so that Li production is dominated by fusion
within Spite plateau metallicities.
On the other hand, spallation is the only mechanism available for cosmic-ray synthesis
of Be and B, and so these have low but measurable abundances at low metallicities.

Our treatment of cosmic rays closely follows that of \cite{FdO3}, so here we only
highlight the essentials.  We assume that supernovae
are the engines of cosmic-ray acceleration, and we therefore scale the cosmic-ray flux
proportionally with the supernova rate.
We assume the cosmic-ray composition
reflects that of the ISM, and thus
set the abundances within cosmic rays to be
that measured today at Earth,
but with CNO scaled by their evolving
ISM abundances.
We assume the energy spectrum remains
the same at acceleration, but that
cosmic-ray escape can very over time.
We thus adopt a source spectrum 
that is a power law in momentum
$\propto p^{-2}$, and an escape column
thickness (``grammage'') that
is 
$\Lambda_{\rm esc} = 20 \ \rm g/cm^2$
at 850 MeV/nucleon, and scales with 
kinetic energy per nucleon as $\epsilon^{-0.6}$.
This escape parameter gives the best \li6/\be9 ratio,
and larger by a factor of $\sim 2$
than that found in present-day cosmic rays, 
suggestive of the ``overconfinement'' scenarios
that have been explored for LiBeB synthesis
in the early Galaxy \cite{Fields:1994kn,crnuke}.

Cosmic-ray interactions with interstellar 
gas produce atoms of species $\ell$
via a variety of reactions.
Cosmic ray nuclei of type $i$ colliding
with interstellar nuclei of type $j$
produce $\ell$ by the process $i + j \rightarrow \ell + \cdots$.
The rate per target $j$ for this process
is $\Phi_i \avg{\sigma_{ij \rightarrow \ell}}$,
where $\Phi_i$ is the cosmic-ray flux of $i$,
$\sigma_{ij \rightarrow \ell}$ is the cross section
for this process, and the brackets
indicate averaging over the cosmic-ray  energy
spectrum.
Thus, the total rate of $\ell$
atom production is the sum over such processes:
$(d{\cal N}_\ell/dt)_{\rm CR} = 
\sum_i \avg{\Phi_i \sigma_{ij\rightarrow \ell}}
{\cal N}_j$,
with the number of interstellar target atoms
is given by ${\cal N}_j=X_j M_{\rm gas}/A_j m_u$, 
and where $A_j$ is the atomic mass of the target and $m_u$ is the atomic mass unit.
The rate of mass production of $\ell$
is given by  $Q_{\ell,\rm CR} \equiv (d{M}_\ell/dt)_{\rm CR} = A_\ell m_u 
(d{\cal N}_\ell/dt)_{\rm CR}$, which is
\begin{equation}
\label{eq:GCRNrate}
    Q_{\ell,\rm CR}  
    \equiv \Gamma_{\ell,\rm CR} M_{\rm gas}
\end{equation}
where the cosmic-ray mass production rate of $\ell$ is
is
\begin{equation}
\label{eq:production}
    \Gamma_{\ell,\rm CR} = \Phi_{p} \sum_{ij} \frac{A_\ell}{A_j} y_i X_j 
    \avg{\sigma_{ij\rightarrow \ell}} \ \ .
\end{equation}
Here $\Phi_{p}$ is the cosmic-ray proton flux,
which we scale with the core-collapse supernova rate
measured as the massive star explosion rate:
$\Phi_{p} \propto R_{\rm CC}$.
Eq.~(\ref{eq:production}) shows that
the mass production rate is weighted by
the cosmic-ray ``projectile'' abundances $y_i=\Phi_i/\Phi_p$,
the ISM ``target'' mass fraction $X_j$,
the relevant mass numbers $A$.

\subsection{Modelling Galactic Chemical Evolution}

We describe chemical evolution in our Galaxy
with the straightforward one-zone model
used in \cite{FdO3}.
This follows the cycling of Galactic 
gas in and out of stars,
given a prescription for star formation,
as well as stellar and cosmic-ray nucleosynthesis yields.
For the star-formation
rate $\psi = dM_\star/dt$,
we choose a form 
proportional to the gas mass:
$\psi = \nu M_{\rm gas}$ with 
$\nu = 0.25 \ \rm Gyr^{-1}$.
For the initial mass function 
$\phi(m) \propto dN_\star/dm$, 
we take the power-law form
$\phi(m) \propto m^{-2.65}$.

If we consider a closed-box model (both open and closed-box models were considered in \cite{FdO3}), the total mass of the Galaxy remains fixed.
The net rate of gas sequestration into stars is
\beq
\label{eq:mgas}
\frac{d}{dt} M_{\rm gas} = - \psi + E 
\eeq
where the stellar gas ejection rate 
$E = \int m_{\rm ej}(m) \ \psi(t-\tau_m) \ \phi(m) \ dm$
is lagged from the birth rate by the lifetime
$\tau_m$ of stars at mass $m$.

For each nucleosynthesis species $i$,
an expression similar to 
eq.~(\ref{eq:mgas}) holds for the mass
$M_{{\rm gas},i}$ in that species,
and the associated mass fraction
$X_i = M_{{\rm gas},i}/M_{\rm gas}$.
For a LiBeB species $\ell$,
we include cosmic-ray production
from eq.~(\ref{eq:GCRNrate}) 
and arrive at
\begin{equation}
\label{eq:massfrac}
\frac{d}{dt} X_\ell 
 =  \frac{E_\ell - X_\ell E}{M_{\rm gas}} 
+ \Gamma_{\ell,\rm CR} 
 =  - \frac{E}{M_{\rm gas}} 
\left( X_\ell - X_\ell^{\rm ej} \right) 
+ \Gamma_{\ell,\rm CR} 
\end{equation}
where $E_\ell = \int m_{{\rm ej,}\ell}(m) \ \psi(t-\tau_m) \ \phi(m) \ dm$ is the rate of mass ejection in
species $\ell$, and where 
$X_{\ell}^{\rm ej} = E_{\ell}/E$
is the mass fraction of $\ell$ in
the ejected gas, and we use the ejection rate $E$ in Eq.~(\ref{eq:mgas}).

We see that the first term in eq.~(\ref{eq:massfrac}) accounts for 
stellar production: it acts to
drive $X_\ell$ to $X_\ell^{\rm ej}$, i.e., 
towards the stellar yields of this species.
These occurs over a timescale $M_{\rm gas}/E \sim M_{\rm gas}/\psi$ that is the characteristic
time for a large portion of
the galaxy's gas to be converted into stars.
Because stars produce `metals' like CNO and Fe, 
their abundances grow
with time.
On the other hand, stars destroy \li6, \be9, and \bor{11}, so for these species $E_\ell = 0 = X_\ell^{\rm ej}$,
and stellar processing leads to their destruction (astration).

The second term in eq.~(\ref{eq:massfrac}), $\Gamma_{\ell,\rm CR}$,
accounts for cosmic-ray nucleosynthesis. 
It is nonzero for LiBeB, while for CNO and Fe this term vanishes.
The net effect for LiBeB is production at early times, which reaches a peak at late times
when the effects astration become important.

Finally, we comment on the normalization of the GCRN rate which should in principle be fixed by the
present cosmic ray total flux. 
Keeping in mind the uncertainty in the flux, we
normalize the cosmic-ray yields so that \be9 attains its solar abundance
at $[{\rm Fe/H}]=0$; this fixes the GCRN contribution for $\li{6,7}$ and $\bor{10,11}$ 
\cite{FdO3,vcfo}. For  \li7 and \iso{B}{11},
there is an additional source from the $\nu$-process
that makes these isotopes via neutrino spallation events in supernovae 
\cite{nuproc}.\footnote{Banerjee et al.~\cite{Banerjee2013} point out the $\nu$-process can produce \be9 via three-step reactions, and that these might be important at early times; these effects are neglected here.} 
Due to the model uncertainties in the $\nu$-process, we scale that contribution \cite{Olive:1993wh,vcfo,vroc}
so that \iso{B}{11}/\iso{B}{10} at
$[{\rm Fe/H}] = 0$ is equal to the observed
ratio, $4.05 \pm 0.16$ \cite{cr}.  
Of course \li7 receives its problematic primordial abundance from BBN,
as well as other sources at higher metallicity.
For completeness we also include the primordial ${\rm \li6/H} \sim 10^{-14}$ abundance,
but this is too low to have any effect on our results.

\begin{figure}[ht!]
    \centering
    \includegraphics[width=0.6\textwidth]{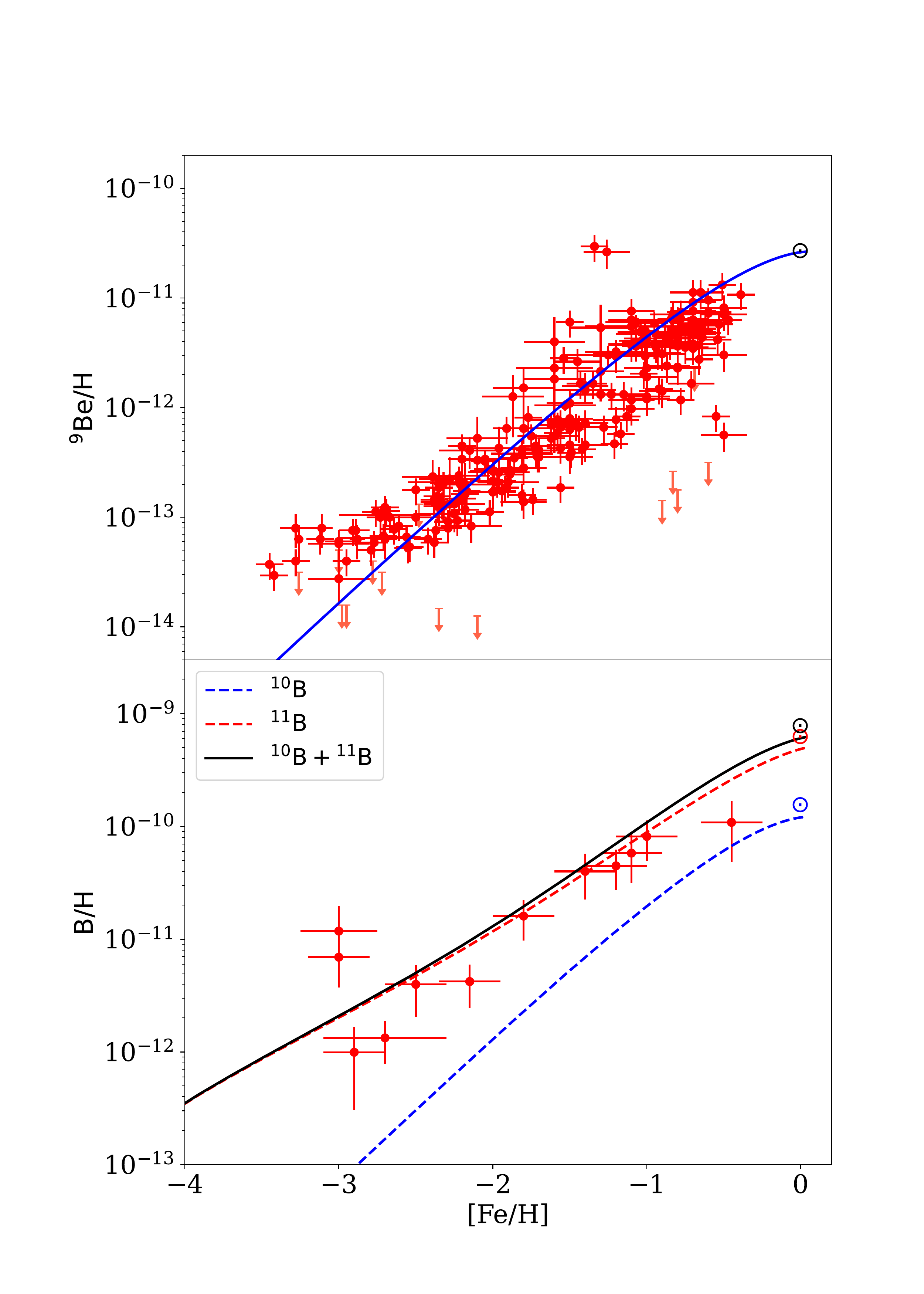}
    \caption{Cosmic-ray production of Be and B evolution vs [Fe/H].  Solar abundances
    are also shown.
    {\em Top panel:} \be9/H; the model is normalized to reproduce the solar value, which fixes all of the cosmic-ray production of LiBeB. Data are taken from \cite{reb,r1,r2,gil,rebnew,bk,pth,ht96,mbe,boes,Primas2000,Smiljanic2009,Spite2019,Smiljanic2021}.  {\em Bottom panel:} the boron isotopes, including \iso{B}{11}
    production from the $\nu$-process. Data are taken from \cite{dbor,gletal,prim}.}
    \label{fig:BeB-Fe}
\end{figure}

\begin{figure}[ht!]
    \centering
    \includegraphics[width=0.6\textwidth]{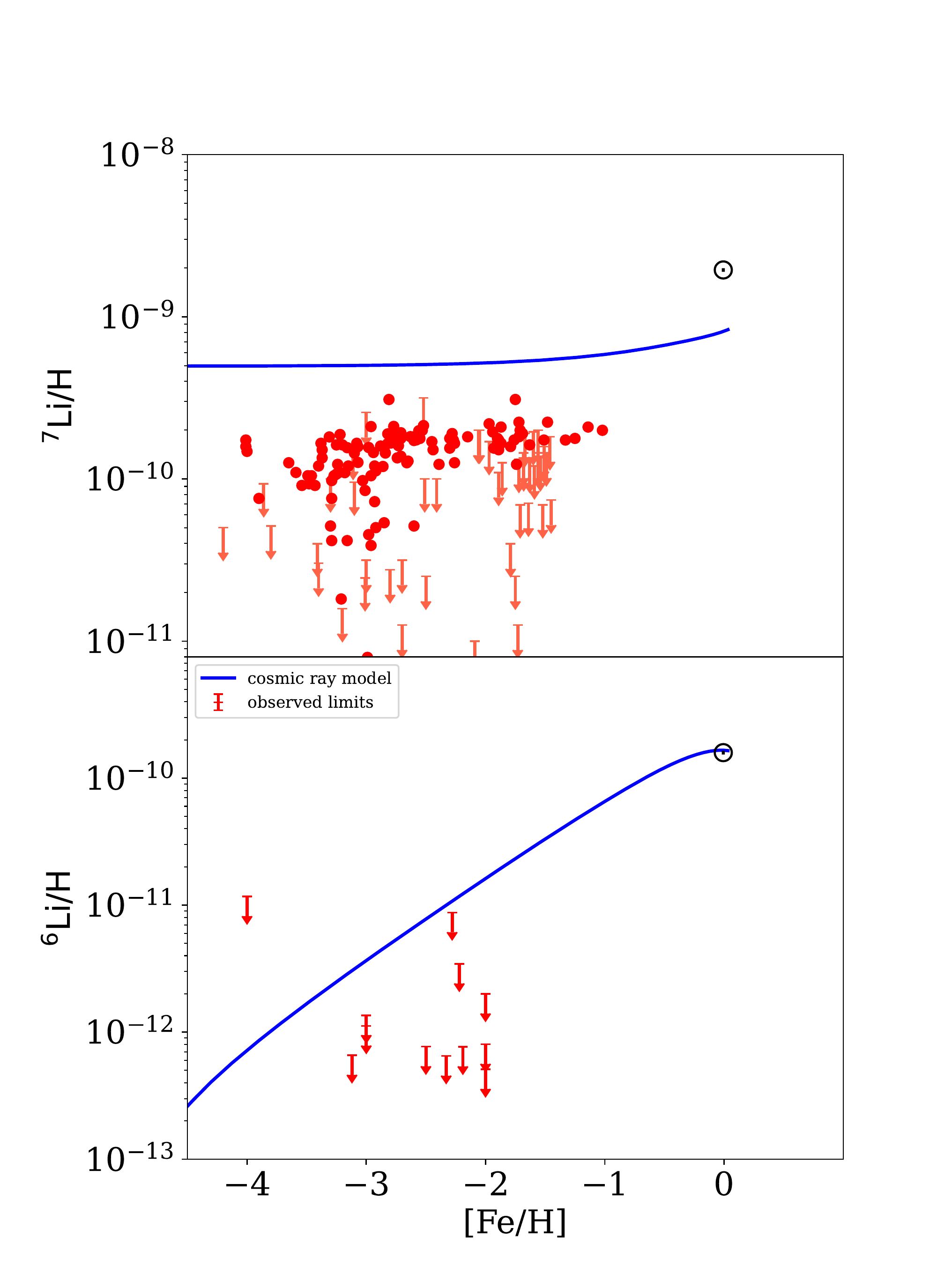}
    \caption{Cosmic-ray prediction for lithium isotope evolution versus [Fe/H].
    Curves show our cosmic-ray model predictions, which include GCR nucleosynthesis
    for both isotopes, and $\nu$-process production for \li7. In addition, for \li7, 
    primordial BBN production is also included.
    {\em Top panel:} The BBN/GCRN \li7/H 
    abundance is contrasted  with elemental
    Li/H abundance
    data compiled from \cite{asplund,bonifacio2007,Melendez:2009xv,aoki2009,Monaco:2010mm,sbordone2010,aoki12,Masseron:2012fh,Bonifacio:2012vp,mucciarelli,Bon15,hansen15,matsuno,Bon18,Aguado:2019egq,rg,gh19,pinto};
    stars shown have $T_{\rm eff}>6000 \ \rm K$. {\em Bottom panel:} The GCRN \li6/H abundance contrasted  with
    data (upper limits) compiled from
    \cite{GP,lind,gh19,wang}.}
    \label{fig:Li-Fe}
\end{figure}

We show in Figs.~\ref{fig:BeB-Fe} and \ref{fig:Li-Fe} the evolution of the LiBeB element
abundances as a function of [Fe/H]. Here, we will restrict our attention to a relatively simple closed-box model.  
As our primary purpose is the implications for Li depletion, we have only normalized to the solar values of Be in Fig.~\ref{fig:BeB-Fe}, and we do not attempt here to fit the evolutionary behavior of these isotopes though as one can see
from the comparison with the data take from \cite{reb,r1,r2,gil,rebnew,bk,pth,ht96,mbe,boes,Primas2000,Smiljanic2009,Spite2019,Smiljanic2021} for Be and \cite{dbor,gletal,prim} for B, the model reproduces the data quite well.\footnote{
The Be data show significant scatter
beyond the observational uncertainties.
This may reflect the origin of halo stars
in accreted dwarf galaxies,
where one would expect a dispersion in the cosmic-ray
and chemical evolution properties
\cite{Pasquini2005,Spite2019}.
This could also introduce dispersion in the
cosmic-ray production of lithium isotopes,
though mitigated by their primary nature.
}

In particular, we do not here consider additional
cosmic-ray sources of the LiBeB elements \cite{clv,rkl,rklr,vroc,superB,Tatischeff2018}
which tend to yield primary BeB, i.e., a linear trend with the spallation target oxygen:
$\log({\rm BeB/H}) = [{\rm O/H}] + \mbox{const}$.  This contrasts with 
standard GCRN which yields secondary BeB
having $\log({\rm BeB/H}) = 2[{\rm O/H}]+\mbox{const}$. 
Observations of BeB as a function of the metallicity tracer [Fe/H]
point to a mixture of primary and secondary components versus iron.  However, as argued in \cite{FdO3,fovc}, a better tracer for the production of BeB is oxygen rather than iron.
If [O/Fe] is constant, then a mixture 
of primary and secondary BeB would require
nucleosynthesis beyond standard GCRN. Data indicate \cite{Israelian:1998fu,boes} that [O/Fe] in fact varies with [Fe/H] leaving open the possibility that the simplest GCRN models can
explain the evolution the BeB isotopes. In any case, this is not of particular concern here, as this issue bears little on the question of the Galactic production of lithium and its depletion in
metal-poor stars,
which is dominated by the primary $\alpha+\alpha$ fusion process.

In Fig.~\ref{fig:Li-Fe}, we show the evolution of the Li isotopes.
At very low metallicity, the evolution of \li7/H is very flat, as it is dominated by
its BBN primordial value. Slowly, as GCRN production (and the $\nu$-process) becomes effective, the abundance of \li7/H begins to rise. Note that we do not 
include any late time production (such as novae \cite{cm}) and thus the present \li7/H abundance falls short of the solar value. In contrast, the primordial \li6/H abundance is very low ($\approx 10^{-14}$) and the GCRN production of \li6 is seen at low metallicity monotonically rising with [Fe/H]. Note that the \li6 vs [Fe/H] results largely follow scaling for a primary process,
so that many of the model details are not essential.
A model parameter that {\em does} have
a significant impact is the $[{\rm O/Fe}]$
slope, which anticorrelates
with the \li6-Fe slope at low metallicity.
We adopt $[{\rm O/Fe}]=\omega_{\rm O/Fe} [{\rm Fe/H}]$, with $\omega_{\rm O/Fe}=-0.35$,
as suggested by observations. 
The \li6 slope then follows
as 
$\log({\rm \li6/H}) = (1+\omega_{\rm O/Fe}) {\rm [Fe/H]}$.
A $\omega_{\rm O/Fe}$ value closer to zero leads to a steeper \li6 slope, and thus less expected
\li6 at low metallicities.

The data for \li6 and \li7 in Fig.~\ref{fig:Li-Fe} come from many sources.
For \li6, we use data giving only upper limits from \cite{GP,lind,gh19,wang}, and for \li7 there has been a substantial amount of new data\footnote{In fact, the data plotted in the top panel of Fig.~\ref{fig:Li-Fe} are elemental lithium abundances that sum the isotopes: ${\rm Li/H}=(\li6+\li7)/{\rm H}$.  The low limits on \li6 demonstrate that in practice the elemental abundances measure \li7. } as discussed above  \cite{asplund,bonifacio2007,Melendez:2009xv,aoki2009,Monaco:2010mm,sbordone2010,aoki12,Masseron:2012fh,Bonifacio:2012vp,mucciarelli,Bon15,hansen15,matsuno,Bon18,Aguado:2019egq,rg,gh19,pinto}.
Because lithium destruction is known to be a sensitive function of effective temperature
(which correlates with the depth of the convective zone),
we follow the standard practice of only showing abundances for hot stars,
with $T_{\rm eff} > 6000 \ \rm K$.

The contrast between predictions and observations in Fig.~\ref{fig:Li-Fe} is illuminating
for both lithium isotopes.
For \li7 we see the familiar deficit in the observed abundances versus the BBN+CMB predicted
primordial value.  
We further see that some recent data indicate a ``meltdown'' of the Spite plateau, implying that
{\em some} depletion has occurred, particularly for stars at very low metallicity.
For \li6, we see that the predicted evolution exceeds most of the observed upper limits,
often by a large factor.  Indeed, 
only {\em one} \li6 limit lies well above 
our predictions!  Here we see the power of the new \li6 limits, which now
imply significant destruction has occurred from the initial cosmic-ray-produced abundances.
This \li6 depletion will be accompanied by \li7 depletion, with consequences for
the primordial Lithium Problem, as we now see.

\section{Consequences of the non-Observation of \li6}
\label{nonobs}

The initial observations \cite{sln,sln2} of \li6 in a few metal-poor stars
with [Fe/H] $\approx -2$, were in good agreement with expectations from cosmic-ray nucleosynthesis \cite{FdO4,Vangioni99}.\footnote{These claimed \li6 detections are not shown in Fig.~\ref{fig:Li-Fe}.} At the same time, these observations
put pressure on stellar models where the depletion of \li6 in particular was expected \cite{pm}. The relative amount of depletion between \li6 and \li7 is model-dependent, but 
\li7 depletion should be no more than that of \li6 due basic considerations
of nuclear physics: \li6 is more weakly bound than \li7
and so its destruction is favored
\cite{BS}.

It is convenient to define the lithium depletion factor as the ratio of the initial stellar abundance to the observed abundance, so that for isotope $i\in (6,7)$: 
\begin{equation}
    D_i \, \equiv \, \frac{{}^{i}{\rm Li_{\rm init}}}{{}^{i}{\rm Li_{\rm obs}}}  \ \ge 1 \  \ .
\end{equation}
With this definition,
the case of no depletion is $D_i = 1$, while
progressively  more depletion means 
progressively larger $D_i$.
In the limit of pure dilution, the depletions are equal: $D_7 = D_6$,
and so a measure of \li6 destruction also gives that of \li7.
On the other hand, nuclear burning depletes \li6 more due to its lower
binding energy, and its smaller mass and hence smaller Coulomb penetration suppression.
Thus, in cases where destruction is incomplete we expect $D_7 < D_6$, and thus \li6 depletion sets an upper limit to \li7 depletion\footnote{However, the pre-main sequence destruction of \li6 could be substantial in contrast to that of \li7.};
this differential depletion has been studied in detailed models \cite{dep6}.

Because \li6 is not detected with certainty in any stars, we can only put a {\em lower}
limit $D_6 > D_6^{\rm lim}$.
That is, $D_6^{\rm lim}$ is determined by the GCRN-predicted
abundance of \li6 at a given value of [Fe/H] relative to the observed value (or upper limit). 
Our approach is to use this limit to estimate the \li7 depletion, by assuming
that 
\beq
\label{eq:Dest}
D_7 = D_6^{\rm lim} \ \  . 
\eeq
In the limit of pure dilution, our lower limits on \li6 depletion
translate to lower limits on the true \li7 destruction:  $D_7 > D_6^{\rm lim}$.
However, in the case of differential depletion, our  \li7 depletion estimate
could be larger or smaller than the true \li7 destruction. Despite this uncertainty,
we will see that the approach embodied in eq.~(\ref{eq:Dest})
gives illuminating results.

We now use our GCRN model to infer \li6 depletion
and to estimate \li7 depletion in metal-poor stars.
For each star with an observed \li6 limit
in Fig.~\ref{fig:Li-Fe}, we compute 
a lower limit to the depletion:
\begin{equation}
D_{6}^{\rm lim} = \frac{\li6_{\rm GCRN}}{\li6_{\rm obs,lim}}    
\end{equation}
We then estimate the \li7 depletion
by  assuming it is the same at our limit for
\li6:  $D_7 = D_6^{\rm lim}$.
We then infer the initial \li7 abundance
in the same star via
\begin{equation}
 \pfrac{\li7}{\rm H}_{\rm init} 
 = D_7 \, \pfrac{\li7}{\rm H}_{\rm obs}
 \approx D_6^{\rm lim} \, \pfrac{\li7}{\rm H}_{\rm obs}
\end{equation}
Thus the larger the \li6 deficit with respect to the GCRN model, the larger the correction to \li7.

\begin{figure}
    \centering
    \includegraphics[width=0.7\textwidth]{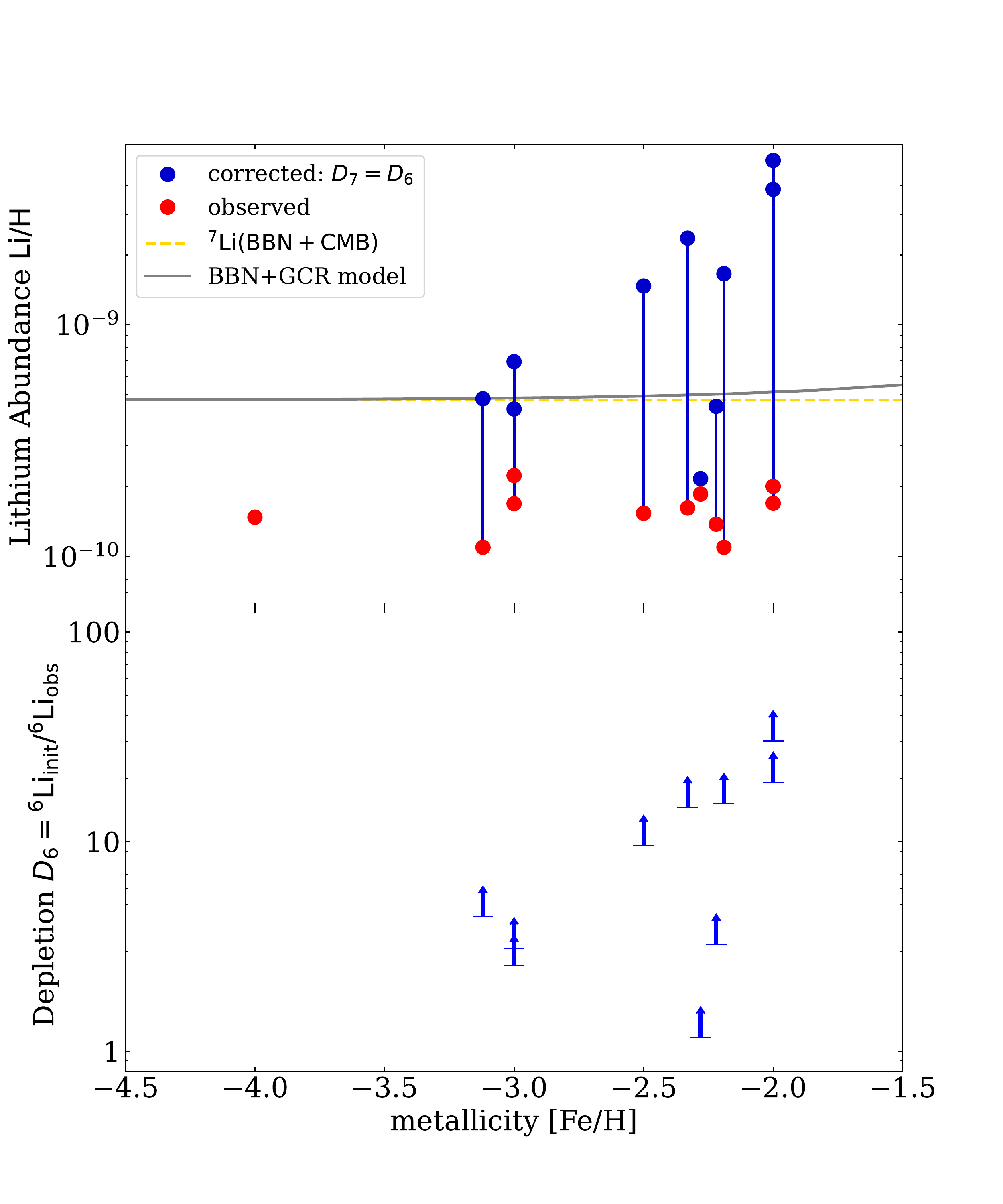}
    \caption{Implications of \li6 depletion.  {\em Bottom Panel:} \li6 depletion factors $D_6 \ge 1$ inferred from 
    the mismatches between observations and model predictions in Fig.~\ref{fig:Li-Fe}. 
    {\em Top Panel:} Elemental lithium abundances $\rm Li/H = (\li6+\li7)/H$.  Points show observed values as well as corrected abundances based on depletions in the lower panel.  Curves are primordial \li7 abundance (eq.~\ref{eq:Lip}) and our model curve for BBN+CMB evolution.}
    \label{fig:LiCorrected}
\end{figure}

Results appear in Fig.~\ref{fig:LiCorrected}.
In the lower panel we see that the depletions
are in many cases very large, up to a factor $> 50$.  This leads to large corrections
to the \li7 abundances, which appear in the upper
panel.
There we see that the corrected \li7/H in most cases easily accommodates
the expected primordial abundance.  

We thus see that the combination of the
strong limits on \li6 in halo stars,
along with cosmic-ray nucleosynthesis, 
together imply that substantial lithium
destruction of {\em both} \li6 and \li7 may have occurred in these stars.
This conclusion is independent of stellar
models for lithium depletion, but
is consistent with and supports many such models.
Indeed, the observed upper limits to the observed \li6 abundance in halo stars, and the implied depletion when comparing these
limits to the abundances predicted in GCRN,
together strengthen the case
that stellar depletion is the culprit in the 
primordial Lithium Problem.  These results 
in turn imply 
that standard BBN is working well and are not in
conflict the observations when similar depletion factors for 
\li6 and \li7 are factored in.

It is important to recall that 
there are is significant model dependencies
to GCRN predictions.  Other reasonable models
would give somewhat different \li6 evolution trends versus metallicity,
and thus lead to different \li6 and \li7 depletions.
It would be useful to explore these uncertainties
thoroughly in future work.
However, we note that the basic primary nature
of \li6 evolution at early times 
sharply  limits the range of possible viable
models for this isotope. Moreover, 
we recall that the \li6 depletions we find are
only {\em lower limits}, and many imply
a \li7 depletion far in excess of what is needed to resolve the Lithium Problem. 
Thus, our main qualitative
result should be robust in that  
substantial lithium depletion is required.

Indeed, we can use our model to 
help discern the nature of 
lithium depletion in halo stars.
We note that for most stars with \li6 limits,
the corrected \li7/H substantially {\em exceeds}
the expected primordial abundance.  The most straightforward interpretation
is that we have overestimated the \li7 destruction, which we assumed to be the same as for
\li6.  That is, fitting \li7 to the required primordial abundance
implies this $D_7 < D_6$.  This indicates differential destruction of the lithium
isotopes, and thus supports stellar evolution
models that have this property.

\section{Discussion and Conclusions}
\label{conc}

Predictions from standard BBN for the \li7 abundance exceed the abundance level seen in Population II stars. This is the crux of the Lithium Problem. In fact, when one folds in the fact that some of the observed lithium is most likely cosmic-ray produced (beryllium and lithium are observed in many of the same stars and beryllium is most certainly produced in cosmic-ray collisions), the \li7 Problem is accentuated \cite{OS}.  

Stellar depletion has always offered a mechanism to solve the Lithium Problem.
But standard stellar models predict \li7 depletion
with significant dispersion \cite{vvsm,duncan81} and the initial observations of the \li7 plateau \cite{spites} put severe constraints on these models \cite{mfb}. Considerable effort went into refining stellar models to limit the amount of dispersion. Recent work \cite{newdep} now attempts to explain
the meltdown at low metallicity.

However the lack of dispersion was not the only
argument against depletion.
The perceived presence \li6 in halo stars reaffirmed the 
primordial nature of \li7. 
Confidence in the early observations was strengthened by the
fact that reported abundance of \li6 matched expectations from simple GCRN models such as that described here. The apparent lack of \li6 depletion implied the lack of \li7 depletion \cite{BS}.

It is now apparent that halo stars not only
show Li dispersion, but also that
\li6 has yet to be definitely observed in any of the hot metal poor halo stars under consideration.
This revised observational outlook
calls of a revised assessment of lithium
depletion.
Indeed to some, it may not be a surprise that 
the \li6 non-observation complements 
the lithium dispersion, both pointing to 
\li7 depletion. 
. 

We have argued that the lack of firm \li6 observations,
with limits far below the levels predicted by GCRN models, implies that \li6 was indeed depleted. 
Since \li6 is only made by cosmic rays, we can relatively  simply model its evolution
to infer the undepleted \li6 at any metallicity, then use the observed abundance to infer
the \li6 depletion.  This in turn provides
and estimate of the \li7 depletion.
An equal amount of \li7 depletion (expected if the primary source of depletion is dilution), exceeds the ratio of
BBN \li7 to the original Spite plateau abundance. Therefore,
not only has the evidence against Li depletion disappeared, 
but its necessity (to explain \li6) may well suggest the absence of 
a primordial \li7 Problem.

Future searches for \li6 remain of prime importance to 
more sharply probe stellar lithium processing as well as cosmic-ray nucleosynthesis.
Additional \li6 limits a can be illuminating. 
Searches for \li6 all along the Spite 
plateau region, will allow for a fuller
quantitative picture of the allowed depletion.
Moreover, if \li6  could be {\em detected} in some stars, then cosmic-ray models will
allow for a {\em measurement} of the \li6 depletion and thereby will probe \li7 depletion as well.
In all cases, the higher metallicity end of the plateau, say $[\rm Fe/H] \sim -2$ to $-1$
offers the highest expected \li6, and thus potentially offer the strongest constraints
on Li destruction.   Even \li6 measurements at still higher metallicity would
be useful to establish the level of depletion in less primitive stars.

In addition to stellar measurements of lithium,
interstellar observations of lithium in extragalactic low-metallicity systems could hold
great promise.  These environments have very different systematics compared to halo stars, most importantly because there is no {\em in situ} lithium destruction in ISM gas.
Elemental lithium has been measured in the ISM of Small Magellanic Cloud, 
at a level similar to that in 
Milky Way disk (Population I) stellar abundances at the same metallicity \cite{Howk12}.
This suggest that there is not large depletion in these stars. But our work here predicts that at lower metallicities, the ISM levels of Li should
be higher than in corresponding halo stars.  Moreover, in interstellar spectra, the Li isotopes can be better distinguished and thus easier to measure 
than in halo stars.  The challenges to overcome for such observations
include finding suitable target systems with sufficiently bright background sources,
and avoiding regions where Li is highly ionized.\footnote{Another extragalactic Li observable is the millimeter-wave signature of the lithium hydride molecule, which has been potentially detected in
one absorption system \cite{fkf};  these are interesting probes of interstellar chemistry but for that reason would be less transparent as measures of lithium nucleosynthesis.}
The payoff would be that such measurements would offer a new probe of lithium evolution, complementary to that of stellar abundances.

In closing, we note that a stellar astrophysics solution 
to the Primordial Lithium Problem
clearly would have profound consequences.
It would 
remove a cloud of lingering concern about standard BBN, and strengthen its role in cosmology and particle physics.
Indeed, nonstandard BBN scenarios have already been increasingly challenged by the high-precision of D/H abundances that agree with the BBN+CMB predictions, and the tight correlation between D and \li7 perturbations in new physics scenarios.
Also, inferring the observed primordial 
lithium abundance would now require the use
of detailed
stellar and cosmic-ray nucleosynthesis models.  So for the near term, \li7 would seem unlikely to be a reliable independent probe 
of BBN--a situation similar to the current status of primordial \he3 determinations \cite{jim}.
It remains to be seen whether future observations can chart a new way to 
measure primordial \li7 unambiguously and
precisely, but we remain ever optimistic.

\section*{Acknowledgments}
We are pleased to thank Tsung-Han Yeh for useful discussions on this paper and related
topics.
We remember our friend, mentor, and collaborator Jim Truran, who taught the authors much of what we know about Galactic chemical evolution in particular and nuclear astrophysics
generally.
The work of K.A.O.~was supported in part by DOE grant DE-SC0011842  at the University of
Minnesota.

\end{document}